\documentclass{article}

\usepackage{amssymb,cite} 
 \usepackage{multirow,slashbox}

\def \BEA { \begin{eqnarray}}
\def \EEA {\end{eqnarray}}
\def \bea { \begin{eqnarray}}
\def \eea {\end{eqnarray}}

\def \BE {\begin{equation}}
\def \EE {\end{equation}}
\def\d{\mathrm{d}}

\def \WDS #1 {\mbox{$\Phi_{#1}^{S}$}}
\def \WDA #1 {\mbox{$\Phi_{#1}^{A}$}}
\def \WD #1 {\mbox{$\Phi_{#1}$}}

\def \mi {\stackrel{i}{m}}
\def \mj {\stackrel{j}{m}}
\def \mk {\stackrel{k}{m}}
\def \mr {\stackrel{r}{m}}
\def \ms {\stackrel{s}{m}}
\def \mp {\stackrel{p}{m}}
\def \mz {\stackrel{z}{m}}
\def \mq {\stackrel{q}{m}}
\def \mo {\stackrel{o}{m}}
\def \mD {\stackrel{2}{m}}
\def \mT {\stackrel{3}{m}}
\def \mC {\stackrel{4}{m}}

\def \mio #1 {\mi_{#1}\ ^{  \! \! \! \! 0}} 
\def \mjo #1 {\mj_{#1}\ ^{  \! \! \! \! 0}} 
\def \mko #1 {\mk_{#1}\ ^{  \! \! \! \! 0}} 
\def \mro #1 {\mr_{#1}\ ^{  \! \! \! \! 0}} 
\def \mso #1 {\ms_{#1}\ ^{  \! \! \! \! 0}} 
\def \mpo #1 {\mp_{#1}\ ^{  \! \! \! \! 0}} 
\def \mzo #1 {\mz_{#1}\ ^{  \! \! \! \! 0}} 
\def \mqo #1 {\mq_{#1}\ ^{  \! \! \! \! 0}} 
\def \moo #1 {\mo_{#1}\ ^{  \! \! \! \! 0}} 
\def \mDo #1 {\mD_{#1}\ ^{  \! \! \! \! 0}} 
\def \mTo #1 {\mT_{#1}\ ^{  \! \! \! \! 0}} 
\def \mCo #1 {\mC_{#1}\ ^{  \! \! \! \! 0}} 

\def \miJ #1 {\mi_{#1}\ ^{  \! \! \! \! (1)}} 
\def \mjJ #1 {\mj_{#1}\ ^{  \! \! \! \! (1)}} 
\def \mkJ #1 {\mk_{#1}\ ^{  \! \! \! \! (1)}} 
\def \mrJ #1 {\mr_{#1}\ ^{  \! \! \! \! (1)}}

\def \bl {\mbox{\boldmath{$\ell$}}}

\def \bn {\mbox{\boldmath{$n$}}}

\def \hbm #1 {\mbox{\boldmath{$\hat m^{(#1)}$}}}
\def \bm {\mbox{\boldmath{$m$}}}

\def \tbl {\mbox{\boldmath{$\tilde{\ell}$}}}
\def \tbn {\mbox{\boldmath{$\tilde{n}$}}}
\def \tbm {\mbox{\boldmath{$\tilde{m}$}}}
\def \tbt {\mbox{\boldmath{$\tilde{t}$}}}
\def \hbt {\mbox{\boldmath{$\hat{t}$}}}

\def \bt {\mbox{\boldmath{${t}$}}}
\def \bT {\mbox{\boldmath{${T}$}}}
\def \bZ {\mbox{\boldmath{${Z}$}}}

\newcommand{\be}{\begin{equation}}
\newcommand{\ee}{\end{equation}}
\newcommand{\beqn}{\begin{eqnarray}}
\newcommand{\eeqn}{\end{eqnarray}}
\newcommand{\pa}{\partial}
\newcommand{\ba}{\begin{array}}
\newcommand{\ea}{\end{array}}
\newcommand{\pp}{{\it pp\,}-}

\def \BEAH {\begin{eqnarray*}}
\def \EEAH {\end{eqnarray*}}
\def \BDM {\begin{displaymath}}
\def \EDM {\end{displaymath}}

\def \pul {{{\footnotesize{\frac{1}{2}}}}}

\newcommand{\tKW}{t}
\newcommand{\yKW}{y}

\newcommand{\FKW}{F}
\newcommand{\GKW}{G}
\newcommand{\accl}{A}                    
\newcommand{\mass}{m}                    

\begin{document}

\title{On higher dimensional Einstein spacetimes with a warped extra dimension}

\author{Marcello Ortaggio\thanks{ortaggio(at)math(dot)cas(dot)cz}, Vojt\v ech Pravda\thanks{pravda@math.cas.cz}, Alena Pravdov\' a\thanks{pravdova@math.cas.cz} \\
Institute of Mathematics, Academy of Sciences of the Czech Republic \\ \v Zitn\' a 25, 115 67 Prague 1, Czech Republic}
\date{\today}

\maketitle

\abstract{
{ We study a class of higher dimensional warped Einstein spacetimes with 
one extra dimension. These were originally identified by Brinkmann as 
those Einstein spacetimes that can be mapped conformally on other 
Einstein spacetimes, and have subsequently appeared in various contexts 
to describe, e.g., different braneworld models or warped black strings.} After clarifying the relation between the general Brinkmann metric and other more specific coordinate systems, we analyze the algebraic type of the Weyl tensor of the solutions. In particular, we describe the relation between Weyl aligned null directions (WANDs) of the lower dimensional Einstein slices and of the full spacetime, which in some cases can be algebraically more special. Possible spacetime singularities introduced by the warp factor are determined via a study of scalar curvature invariants and of Weyl components measured by geodetic observers. Finally, we illustrate how Brinkmann's metric can be employed to generate new solutions by presenting the metric of spinning and accelerating black strings in five dimensional anti-de~Sitter space.}

\section{Introduction}

In recent years there has been a growing interest in gravity in $n>4$ dimensions, mainly motivated by modern unified theories (such as string theory), AdS/CFT and recent brane world scenarios. In particular, in the context of the study and classification of exact solutions, an $n>4$ generalization of the Petrov classification \cite{Coleyetal04}, of the Newman-Penrose (NP) equations \cite{Pravdaetal04,Coleyetal04vsi,OrtPraPra07} and of the Geroch-Held-Penrose  (GHP) formalism \cite{Durkeeetal10} have been  presented, and recently employed in several studies. Many of these have focused on properties of Einstein spacetimes (defined by $R_{ab}=Rg_{ab}/n$), as a natural first step towards the understanding of higher dimensional gravity. Einstein spacetimes indeed describe systems in which there is no matter and only gravity is at play, at the same time allowing for a cosmological constant. They include, e.g., a number of black hole, black string and black ring solutions (see, e.g., \cite{EmpRea08} for a review and references), and they are thus an interesting arena for testing and applying the general methods mentioned above.

In this context, it is the purpose of the present paper to analyze algebraic and optical properties as well as singularities of a special class of higher dimensional Einstein spacetimes, namely those that can be mapped conformally (and ``properly'') on other Einstein spaces. These were fully classified by Brinkmann already in the 1920s \cite{Brinkmann25}, and their  line element takes a specific warped form with a single ``extra dimension''. 
However, a discussion in terms of their possible Weyl type and of the recently developed NP (or GHP) formalism has not been performed yet (but see~\cite{PraPraOrt07} for a discussion of other properties of warped spacetimes). From our perspective, the interest in such a study is twofold. On the one hand, we shall clarify several features of such a general class of spacetimes, and relate these to certain previously known solutions (mentioned later in the appropriate context). On the other hand, from a more pragmatical viewpoint, we shall illustrate how the Brinkmann line element can be used to generate Einstein spacetimes with given algebraic properties and optics. It is well known how difficult it may be, in general, to find exact solutions to the Einstein equations. This simple method can thus prove useful, e.g., in constructing explicit examples (or ``counterexamples'') to test or falsify certain properties of higher dimensional gravity that one might conjecture to hold on the basis of known results in four dimensions. For example, the Brinkmann ansatz  has been already applied in the context of (the geodetic part of) the Goldberg-Sachs theorem
\cite{PraPraOrt07} (cf.~also \cite{Ortaggio07}), and to generate specific examples of Robinson-Trautman \cite{Ortaggio07} and type III/N \cite{OrtPraPra10} Einstein spacetimes (see, respectively, \cite{PodOrt06} and \cite{Pravdaetal04} for general properties of such families of solutions in higher dimensions). It will thus be useful to understand in more generality what kind of spacetimes one may generate with that method. 

Furthermore, we observe that the Brinkmann line element is essentially a slicing of an Einstein spacetime by hypersurfaces which are, in turn, also Einstein. For this reason, it is of interest in brane-world scenarios, where it provides a consistent embedding of $(n-1)$-dimensional Einstein gravity in $n$-dimensional Einstein gravity (with various possible values for the bulk and lower dimensional cosmological constants, see subsection~\ref{subsec_alternative} below). For example, the well known warped metric for the ground state of the Randall-Sundrum (RS) models \cite{RanSun99a,RanSun99b} is in fact a special instance of Brinkmann's spacetimes. More general Brinkmann metrics have been considered in the context of other braneworld Kaluza-Klein (KK) reductions, see, e.g., \cite{LuPop01,CveLuPop01,ParPopSad02} and references therein, where various supergravity extensions (relying on the same metric ansatz) have also been studied. Although we will not be directly concerned with these models here, the results of the present work can thus be of interest also in a wider context. 
The paper is organized as follows.  

In Section~\ref{sec_metric} we present the Brinkmann metric and determine the allowed combinations of signs of the cosmological constants of the  $(n-1)$-dimensional line element $\d\tilde s^2$ and of the $n$-dimensional warped line element $\d s^2$. We also determine coordinate transformations to various metric forms (suitable in different specific contexts) that are equivalent to the Brinkmann metric for appropriate choices of parameters.

In Section~\ref{sec_weyl} the connection between the Weyl types of the metric $\d\tilde s^2$ and  $\d s^2$ is studied. It is shown that $\d s^2$ inherits WANDs from $\d\tilde s^2$ with the same multiplicity. In particular cases, $\d s^2$ can however also possess additional WANDs unrelated to those (if any) of $\d\tilde s^2$. Consequently, $\d s^2$ is {in general} of the same Weyl type of $\d\tilde s^2$, but in particular cases it can be more special. 
 
Using scalar curvature invariants and components of the Weyl tensor in a parallelly propagated (p.p.) frame, in Sections~\ref{subsec_invariants} and \ref{sec_ppsing} we study curvature singularities arising {in the full spacetime $\d s^2$ due to the warp factor}. Such singularities appear in all cases except when both the cosmological constant of $\d\tilde s^2$ and that of $\d s^2$ are negative, and in the trivial case of a direct product spacetime.  

In Section~\ref{sec_examples} we discuss two explicit {five-dimensional} examples of warped metrics {\em without} naked singularities  -- an anti-de~Sitter (AdS) black string sliced by an AdS spinning black hole, and an accelerated AdS black string generated from the four-dimensional AdS C-metric -- {and indicate how to easily generate more general solutions.}

After brief concluding remarks in Section~\ref{sec_conclusions}, we give some necessary technical details about the general warped metric (such as the Christoffel symbols and the components of the Weyl tensor in a parallelly propagated frame, etc.) in Appendices A and B.

\paragraph{Notation} 

Following \cite{Coleyetal04,Pravdaetal04,Coleyetal04vsi,OrtPraPra07,Durkeeetal10}, we use a frame consisting of two null vectors $\bm_{(0)}=\bl$ and $\bm_{(1)}=\bn$, and $n-2$ orthonormal spacelike vectors $\bm_{(i)}$,  
where $i, j, \dots=2,\ldots,n-1$. In terms of these, the metric reads
\be
 g_{ab}=2l_{(a}n_{b)}+\delta_{ij}m_{(i)a}m_{(j)b} , \label{metricIII}
\ee 
where, hereafter, $a,\ b=0,\ 1,\dots,\ n-1$.

In the following, one of the $\bm_{(i)}$ will be naturally singled out because of the metric ansatz. We shall denote this by $\bm_{(Z)}$ and the remaining spacelike vectors of the basis by $\bm_{(I)}$, with $I=2,\ldots,n-2$.

The optical matrix $L$ of $\bl$ is defined by its matrix elements
\be
 L_{ij}=\ell_{a;b}m_{(i)}^am_{(j)}^b , 
\ee  
with (anti-)symmetric parts 
\be
 S_{ij}=L_{(ij)} , \qquad A_{ij}=L_{[ij]} . 
\ee
The optical scalars expansion, $\theta $, shear, $\sigma$, and twist, $\omega$, are defined by $\theta =L_{ii}/(n-2)$,
$\sigma^2=(S_{ij}-\theta \delta_{ij})(S_{ij}-\theta \delta_{ij})$, and $\omega^2=A_{ij}A_{ij}$, respectively. 

For spacetimes of Weyl type III and N we introduce the compact symbols 
\be
  \Psi_{i} = C_{101i}, \qquad  \Psi_{ijk}= \frac{1}{2} C_{1kij}, \qquad \Psi_{ij} = \frac{1}{2} C_{1i1j} ,
\ee
which obey the following constraints  \cite{Pravdaetal04} 
\be 
 \Psi_i=2 \Psi_{ijj}, \qquad  \Psi_{ijk}+\Psi_{kij}+\Psi_{jki}=0, 
 \qquad  \Psi_{ijk}=-\Psi_{jik}, \qquad  \Psi_{ij}=\Psi_{ji}, \qquad  \Psi_{ii}=0.
\ee

\section{General metric form}

\label{sec_metric}

\subsection{Brinkmann coordinates}

\label{subsec_brinkmann}

We study {$n$-dimensional} warp product metrics of the form 
\be
 \d s^2=\frac{1}{f(z)}\d z^2+f(z)\d\tilde s^2 ,  
 \label{ansatz} 
\ee
{where $\d\tilde s^2$ is an $(n-1)$-dimensional metric.} 
Assuming that d$s^2$ is an Einstein spacetime (i.e. $R_{ab}=Rg_{ab}/n$ and the Ricci scalar $R$ is related to the cosmological constant by $(n-2)R=2n\Lambda$) it follows that (see  Appendix \ref{app_riemann})
\be
 f(z)=-\lambda z^2+2dz+b , \qquad \lambda=\frac{2\Lambda}{(n-1)(n-2)} ,
 \label{fz}
\ee
with $b$ and $d$ being constant parameters. Then $\d\tilde s^2$ { turns out to be} also Einstein, with Ricci scalar (hereafter tildes will denote quantities referring to the geometry of $\d\tilde s^2$)\footnote{From the braneworld KK reduction viewpoint, $\tilde R$ corresponds to the cosmological constant in the lower dimensional spacetime.}
\be
 \tilde R=(n-1)(n-2)(\lambda b+d^2).
\label{ricci-n-1}
\ee 

In his early work \cite{Brinkmann25}, Brinkmann showed that an Einstein spacetime can be mapped conformally on another Einstein spacetime by a {\em proper} map\footnote{A conformal map $\hat g_{ab}=\Omega^2g_{ab}$ is called proper if $g^{ab}\Omega_{,a}\Omega_{,b}\neq0$ \cite{Brinkmann25}. Improper conformal maps are possible only between Ricci-flat spacetimes, which in fact must be \pp waves (with a Ricci-flat ``transverse'' space) \cite{Brinkmann25}. Since these have been already thoroughly investigated (see, e.g., \cite{Coleyetal03,Coleyetal04vsi,Coleyetal06}), we will restrict to proper maps in this paper.}
if and only if its line element can be written in the form  (\ref{ansatz}) with (\ref{fz}) (cf., e.g., also \cite{petrov}). {This invariantly characterizes the class of considered metrics, which have also} appeared in {other} contexts to describe, e.g., different braneworld models or warped black strings. {For example,} one of the Einstein metrics conformal to $\d s^2$ is given by $\d\hat s^2=z^{-2}\d s^2$ \cite{Brinkmann25}. Of course, it must be possible to put also $\d\hat s^2$ in the form~(\ref{ansatz}), as one can indeed do by defining a new coordinate $\hat z=1/z$ and taking $\hat f(\hat z)=b\hat z^2+2d\hat z-\lambda$. Since $R=n(n-1)\lambda$ one immediately gets $\hat R=-n(n-1)b$. {See \cite{Brinkmann25,petrov} for further details.}

{The approach of this paper will be to consider} the line element~(\ref{ansatz}) as a useful ansatz to generate an $n$-dimensional Einstein spacetime $\d s^2$ from a known $(n-1)$-dimensional one, i.e. $\d\tilde s^2$. We will refer to $\d\tilde s^2$ as the ``seed'' metric, or as a ``section'' {or ``slice''} of $\d s^2$.  The metric $\d s^2$ is clearly warped, with a warp factor $f(z)$ depending only on the single extra dimension $z$. One obtains in particular a direct product space in the special case of a constant $f(z)$, i.e. $\lambda=0=d$ (with $b>0$). {Although Brinkmann's work \cite{Brinkmann25} is essentially signature independent, here we are interested in studying (Einstein) spacetimes, i.e. we shall assume a Lorentzian signature for $\d s^2$. Moreover, we will restrict to the case in which $z$ is a space coordinate in the ansatz~(\ref{ansatz}), so that $\d\tilde s^2$ must be an Einstein spacetime.\footnote{One can equally consider~(\ref{ansatz}) in the case of a timelike $z$. These metrics are simply obtained from (\ref{ansatz}) by Wick rotating $z=it$ along with $d=id'$ and $b=-b'$, so that $\d s^2=-f^{-1}(t)\d t^2+f(t)\d\tilde s^2$, where $f(t)=\lambda t^2-2d't-b'$, and $\d\tilde s^2$ must be an Euclidean Einstein space with Ricci scalar $\tilde R=-(n-1)(n-2)(\lambda b'+d'^2)$.  Such solutions are of course time-dependent. Some of their properties  will straightforwardly follow from the results of the present paper. On the other hand, one cannot in general extend statements about WANDs and the Weyl type, since real WANDs may become complex after a Wick rotation.
In a wider context, however, several results about general warped spacetimes with a one-dimensional Lorentzian factor (which include, in particular, all static spacetimes) are known. For example, their Weyl type can be only G, I$_i$, D or O \cite{PraPraOrt07}, for all of which we know explicit vacuum examples: static KK bubble (of type G \cite{GodRea09}), static black ring (of type I$_i$ \cite{PraPra05}) and Schwarzschild black hole (of type D \cite{Coleyetal04,Pravdaetal04,PraPraOrt07}). We will thus not be concerned with such a discussion here.}
We thus} require $f(z)>0$, which may restrict possible parameter values and (possibly) the range of $z$. Namely, {since $f(z)$ has real roots if and only if $\tilde R \geq 0$},
when $\tilde R \le 0$ we require $\lambda<0$ ($\tilde R = 0$ admits also $\lambda=0$, but this case simply corresponds to a direct product), while for $\tilde R>0$ any sign of $\lambda$ (including $\lambda=0$) is admitted, at least for suitable values of $z$.

We finally observe that (\ref{ansatz}) is form-invariant under a redefinition $z=\alpha z'+\beta$, under which $\lambda'=\lambda$, $d'=\alpha^{-1}(d-\lambda\beta)$,  $b'=\alpha^{-2}[b+\beta(-\lambda\beta+2d)]$, and $\d\tilde s'^2=\alpha^{2}\d\tilde s^2$ (so that $\tilde R'=\alpha^{-2}\tilde R$). This freedom will be useful later on.

\subsection{Alternative forms of the metric}

\label{subsec_alternative}

The coordinate system employed so far has the advantage that allows for a unified treatment of all metrics of the form~(\ref{ansatz}), regardless of the specific value of the parameters $\lambda$, $d$ and $b$ entering $f(z)$. However, for certain applications other modified coordinates may sometimes be more convenient.

Before illustrating those, let us first reduce the number of parameters of~(\ref{ansatz}). Indeed, by an appropriate redefinition of $z$ (and, possibly, a rescaling of $\d\tilde s^2$) one can always rewrite the line element~(\ref{ansatz}) in the following ``normalized'' forms 
\be
 \d s^2=\frac{1}{-\lambda z^2+\epsilon}\d z^2+(-\lambda z^2+\epsilon)\d\tilde s^2  \qquad (\epsilon=\pm1,0) , \quad \mbox{with } \tilde R=(n-1)(n-2)\lambda\epsilon ,
 \label{ansatz2} 
\ee
where, in order to have a correct signature, the values $\epsilon=-1,0$ require $\lambda<0$. In the case $\lambda=0$ the additional metric is possible
\be
 \d s^2=\frac{1}{2z}\d z^2+2z\d\tilde s^2 \qquad (\lambda=0), \quad \mbox{with } \tilde R=(n-1)(n-2) .
 \label{ansatz0+} 
\ee

We have thus six different, inequivalent metrics corresponding to different choices of the parameters in the original ansatz~(\ref{ansatz}), and fully characterized by the signs of the Ricci scalars $R$ (or $\lambda$) and $\tilde R$ of the full spacetime $\d s^2$ and of its Einstein section $\d\tilde s^2$, respectively. These can be thus schematically summarized as: $(+,+)$, $(-,+)$, $(-,0)$, $(-,-)$, $(0,0)$ (all contained in~(\ref{ansatz2})), and $(0,+)$ (given by~(\ref{ansatz0+})), see Table~\ref{table_combinations}. In particular, note that a negative cosmological constant allows for Einstein sections with any sign of $\tilde R$. As we shall discuss in Sections~\ref{subsec_invariants} and \ref{sec_ppsing}, the above warping in general produces spacetime singularities, except in the cases $(-,-)$ and  $(0,0)$.

\begin{table}[h]
\begin{center}
\begin{tabular}{|c||c|c|c|} 
  \hline 
  \backslashbox {$\tilde R$}{$R$} & $-$  & $0$ &  {$+$}  \\ \hline\hline   
  $-$ & {\footnotesize{$\surd$}}  \quad [(\ref{--a}), (\ref{--b})] & $\times$ \quad & $\times$ \\ \hline 
  $0$ & {\footnotesize{$\surd$}} \quad [(\ref{-0a}), (\ref{-0b})] & {\footnotesize{$\surd$}} \quad [(\ref{00a}), (\ref{00b})] & $\times$  \\ \hline 
  $+$ & {\footnotesize{$\surd$}} \quad [(\ref{-+a}), (\ref{-+b})]  & {\footnotesize{$\surd$}} \quad [(\ref{0+a}), (\ref{0+b})] & {\footnotesize{$\surd$}} \quad [(\ref{++a}), (\ref{++b})] \\ \hline 
\end{tabular} 
\end{center}
		\caption{Allowed choices of signs of the Ricci scalars $R$ {(or $\lambda$)} and $\tilde R$ for the metric (\ref{ansatz}) are denoted by the symbol {\footnotesize{$\surd$}} {(and forbidden ones by $\times$). The corresponding line elements in the alternative coordinates of Sections~\ref{subsubsec_KKcoords} and \ref{subsubsec_conformal_coords} are also indicated (in square brackets).}}
 \label{table_combinations}
\end{table}

\subsubsection{``Braneworld Kaluza-Klein reduction'' coordinates}

\label{subsubsec_KKcoords}

The first natural choice is to replace $z$ by a new coordinate $y$ such that the metric component along the extra dimension is a constant (normalized to 1), i.e. $\d z^2/f(z)=\d y^2$. This leads to
\beqn
 \lambda>0: \qquad & \d s^2 & =\d y^2+\cos^2(\sqrt{\lambda}y)\d\tilde s^2  \qquad\   (\tilde R>0) , \label{++a} \\
 \lambda=0: \qquad & \d s^2 & =\d y^2+\d\tilde s^2 \qquad\qquad \qquad  \ \ \ (\tilde R=0) , \label{00a} \\
 									 & \d s^2 & =\d y^2+y^2\d\tilde s^2 \qquad\qquad\qquad  (\tilde R>0) , \label{0+a} \\
 \lambda<0: \qquad & \d s^2 & =\d y^2+\cosh^2(\sqrt{-\lambda}y)\d\tilde s^2 \qquad (\tilde R<0) , \label{--a} \\
 						& \d s^2 & =\d y^2+e^{2\sqrt{-\lambda}y}\d\tilde s^2 \qquad\qquad \ \ \ (\tilde R=0) , \label{-0a} \\
 						& \d s^2 & =\d y^2+\sinh^2(\sqrt{-\lambda}y)\d\tilde s^2 \qquad\  (\tilde R>0) . \label{-+a}				  							 
\eeqn
We have indicated  the sign of $\tilde R$ above. Its modulus is given by $|\tilde R|=(n-1)(n-2)|\lambda|$, except for the metric~(\ref{0+a}), which has  $\tilde R=(n-1)(n-2)$. As remarked in \cite{ParPopSad02}, this (as well as the next) coordinate system hides the fact that the lower dimensional cosmological constant (or $\tilde R$) is in fact an adjustable parameter. To see this, one can use the coordinate transformation given in \cite{ParPopSad02}, or simply retain the original metric~(\ref{ansatz}) (see the comment concluding subsection~\ref{subsec_brinkmann}). 

It is worth observing that the above metrics have already separately been considered  in the literature from different viewpoints. For example, see \cite{EmpJohMye99} for an application of the AdS metrics (\ref{--a})--(\ref{-+a}) in the AdS/CFT correspondence. The line element (\ref{--a}) appeared also in \cite{ParPopSad02} in the context of the KK reduction of gauged supergravities in $n$ dimensions to gauged supergravities in $n-1$ dimensions. Similarly, the metrics (\ref{++a}) and (\ref{-+a}) were used for embedding certain gauged dS supergravities into gauged dS and AdS supergravities, respectively \cite{ParPopSad02}. Note also that the line element (\ref{-0a}) with a flat $\d\tilde s^2$ is the well known metric of the RS models \cite{RanSun99a,RanSun99b} (except that (\ref{-0a}) does not contain any brane at $y=0$) and, with a generic Ricci-flat $\d\tilde s^2$, has been considered, e.g., in \cite{LuPop01,CveLuPop01,ParPopSad02}. Metric~(\ref{0+a}) was employed, for instance, in \cite{MasLiuWes94} (in five dimensions). 
Specific metrics of the form (\ref{++a})--(\ref{-+a}) appeared in \cite{GodRea09} to describe various black strings.
Very recently metrics  (\ref{++a})--(\ref{-0a}) were also discussed in \cite{Yang10}.

\subsubsection{``Conformal to a direct product'' coordinates}

\label{subsubsec_conformal_coords}

Another natural coordinate system can be constructed such that $\d s^2$ becomes manifestly conformal to a direct product, i.e., $\d z^2/f^2(z)=\d x^2$. One thus finds the possible metrics (presented in the same order as the ones above)
\beqn
 \lambda>0: \qquad & \d s^2 & =\cosh^{-2}(\sqrt{\lambda}x)(\d x^2+\d\tilde s^2) \qquad (\tilde R>0) , \label{++b} \\
 \lambda=0: \qquad & \d s^2 & =\d x^2+\d\tilde s^2 \qquad\qquad\qquad\qquad \ \   (\tilde R=0) , \label{00b} \\
 									 & \d s^2 & = 2e^{2x}(\d x^2+\d\tilde s^2)\qquad\qquad \qquad (\tilde R>0) , \label{0+b} \\
 \lambda<0: \qquad & \d s^2 & =\cos^{-2}(\sqrt{-\lambda}x)(\d x^2+\d\tilde s^2) \qquad (\tilde R<0) , \label{--b} \\
 						& \d s^2 & =(-\lambda x^2)^{-1}(\d x^2+\d\tilde s^2) \qquad\qquad (\tilde R=0) , \label{-0b} \\
 						& \d s^2 & =\sinh^{-2}(\sqrt{-\lambda}x)(\d x^2+\d\tilde s^2) \qquad (\tilde R>0) .	\label{-+b}	 									  									 
\eeqn
Obviously (\ref{00a}) trivially coincide with (\ref{00b}), and with (\ref{ansatz2}) with $\lambda=0$, $\epsilon=+1$, but we have repeated them for completeness. 

{Some of the above metrics have been used to construct various (non-uniform) AdS black strings in five dimensions. Namely, with (\ref{-0b}) Ref.~\cite{ChaHawRea00} studied an AdS black string giving rise to a Schwarzschild black hole on a brane in the RS scenarios. Similarly, AdS black strings foliated by Schwarzschild-(A)dS black holes, relying on (\ref{--b}) and (\ref{-+b}), appeared in \cite{HirKan01}. All such solutions straightforwardly extend to any higher dimensions \cite{GodRea09} (but most of them contain naked singularities, cf. Sections~\ref{subsec_invariants} and \ref{sec_ppsing} and the given references). Obviously, in the case $\lambda=0=\tilde R$, eq.~(\ref{00b}) includes the metric form of the uniform (direct product) Schwarzschild and Kerr black strings in any dimensions.}


\section{Weyl type}
\label{sec_weyl}

We now give the Weyl components for the spacetime~(\ref{ansatz}) and study its possible algebraic type.

\subsection{Weyl frame components}

Using coordinates $x^a=(x^\mu,z)$ {(with Greek indices ranging from 0 to $n-2$)}, for the coordinate components of the Weyl tensor it is straightforward to show that \cite{Brinkmann25} (see also Appendix~\ref{app_riemann}) 
\be
 C_{\mu\nu\rho\sigma}=f\tilde C_{\mu\nu\rho\sigma} , \qquad C_{z\mu\nu\rho}=0=C_{z\mu z\nu} .
 \label{weyl}
\ee

First, it is obvious that $\d s^2$ is conformally flat (and thus of constant curvature since Einstein) iff $\d\tilde s^2$ is such \cite{Brinkmann25}. It thus follows that in four dimensions the metric~(\ref{ansatz}) describes only spaces of constant curvature, since a three dimensional Einstein space $\d\tilde s^2$ is necessarily of constant curvature. However, this is not the case for $n>4$. 
 Let us now assume we are given a null frame in the spacetime $\d\tilde s^2$, consisting of the vectors $\tbl$, $\tbn$ and $\tbm_{(I)}$ {(from now on $I,J=2,\ldots,n-2$)}. We can straightforwardly lift this to a null frame of $\d s^2$ by simply taking
\be
 \bl=\tilde\ell^\mu\pa_\mu, \qquad \bn=f^{-1}\tilde n^\mu\pa_\mu, \qquad \bm_{(I)}=f^{-1/2}\tilde m^\mu_{(I)}\pa_\mu, \qquad \bm_{(Z)}=\sqrt{f}\pa_z , \label{lifted_frame}
\ee  
so that only $\bm_{(Z)}$ will have a nonzero $z$ component.

Then it is easy to see that the only nonzero independent  Weyl frame components, ordered by boost weight (b.w.), are given by
\beqn
 & & C_{0I0J}=\tilde C_{0I0J}, \qquad C_{0IJK}=f^{-1/2}\tilde C_{0IJK}, \nonumber \\ 
 & & C_{01IJ}=f^{-1}\tilde C_{01IJ} , \qquad C_{IJKL}=f^{-1}\tilde C_{IJKL} , \label{weyl_coord} \\
 & & C_{1IJK}=f^{-3/2}\tilde C_{1IJK}, \qquad C_{1I1J}=f^{-2}\tilde C_{1I1J} , \nonumber 
\eeqn
where (from now on) the components of the [un]tilded Weyl tensor are evaluated in the [un]tilded frame.

\subsection{A WAND of the seed metric lifts to a WAND of the full space (with the same multiplicity)}
\label{sec_sharedwand}

From~(\ref{weyl_coord}) it immediately follows that {\em if $\tbl$ is a WAND of $\tilde C_{\mu\nu\rho\sigma}$, then $\bl$ (as defined in~(\ref{lifted_frame})) is automatically a WAND of $C_{abcd}$, with the same multiplicity of $\tbl$}. Therefore, in particular, {\em the Weyl type of the geometry $\d s^2$ is at least as special as the Weyl type of the geometry $\d\tilde s^2$}. {Let us now discuss various possibilities separately.}

If $\d\tilde s^2$ is of type N then obviously (from~(\ref{weyl_coord})) $\d s^2$ is also of type N, and $\bl$ (as in~(\ref{lifted_frame})) is the only WAND. If $\d\tilde s^2$ is of type III then $\d s^2$ is also of type III: if it were of type N then it would admit two distinct WANDs $\bl$ and $\bl'$ of multiplicity, respectively, 3 and 4, which is not possible (it would imply that the Weyl tensor vanishes).
By a similar argument, if $\d\tilde s^2$ is of type II then $\d s^2$ is also of type II,
since type III[N] will lead to having two distinct WANDs of multiplicity 2 and 3[4], which again is forbidden given that $C_{abcd}\neq0$. Thus, {\em if the Weyl tensor of $\d\tilde s^2$ is of type II or more special, then $\d s^2$ is of the same principal Weyl type}. {Note in addition that if $\d\tilde s^2$ is of (secondary) type D (i.e., it possesses at least two distinct double WANDs) then $\d s^2$ is also of type D, since both WANDs can be lifted.}

This argument, however, does not work when $\d\tilde s^2$ is of type I or G, i.e., $\d s^2$ can be more special {in those cases}. We now work out under what conditions there may exist WANDs of $\d s^2$ that are inherently ``higher dimensional'', i.e. not obtainable by simply lifting a WAND of $\d\tilde s^2$, {and  we single out cases in which $\d s^2$ may be more special than $\d\tilde s^2$}. 

\subsection{A WAND of the full space need not correspond to a WAND of the seed metric}

Let us first study under what conditions a new {\em single} (i.e., not multiple) WAND arises in $\d s^2$. We thus assume we have a null vector $\bl$ in $\d s^2$, with $\ell^z\neq 0$. By the Bel-Debever criteria \cite{Milsonetal05,Ortaggio09} this is a WAND if and only if the timelike vector $\tilde {\mbox{\boldmath{$t$}}}=\ell^\mu\pa_\mu$ in $\d\tilde s^2$ satifies 
\be
 \tilde C_{\mu\nu\rho\sigma}\tilde t^\nu\tilde t^\sigma=0 .
 \label{1WAND}
\ee
This constraints the possible geometries $\d\tilde s^2$ but, {\em per se}, does not lead to any general conclusions about existence and multiplicity of WANDs of $\d\tilde s^2$. It is, in particular, compatible with its Weyl type being G or I.\footnote{In the next paragraph we will give an explicit example of a type G spacetime satisfying the stronger condition~(\ref{2WAND}), and {\em a fortiori} also~(\ref{1WAND}).}
When~(\ref{1WAND}) is satisfied, the vector $\bl'=\ell^\mu\pa_\mu-\ell^z\pa_z$ (which is clearly null thanks to $g_{\mu z}=0$) is also a WAND, so that the Weyl type is I$_i$ (or more special, if there is a multiple WAND shared by $\d s^2$ and $\d\tilde s^2$, as discussed  in Section~\ref{sec_sharedwand}), even if the seed metric $\d\tilde s^2$ was of type G. Conversely, given any $\d\tilde s^2$ admitting a timelike vector $\tilde {\mbox{\boldmath{$t$}}}$ that satisfies~(\ref{1WAND}), one can always construct two WANDs $\ell^\mu\pa_\mu\pm\ell^z\pa_z$ of $\d s^2$ (where $\ell^z\neq0$ is fixed by requiring $\ell^\mu\pa_\mu\pm\ell^z\pa_z$ to be null).

If instead $\bl$ is a {\em double} WAND in $\d s^2$ (with $\ell^z\neq 0$), the Bel-Debever criteria \cite{Ortaggio09} give the equivalent condition on the timelike vector $\tilde t=\ell^\mu\pa_\mu$ in $\d\tilde s^2$
\be
 \tilde C_{\mu\nu\rho\sigma}\tilde t^\sigma=0 .
 \label{2WAND}
\ee
This implies, in particular, that $\d\tilde s^2$ can only be of the types G, I$_i$ or D (or O, in which case, however, $\d s^2$ is also of type O) \cite{Ortaggio09}.\footnote{More precisely, the only possible type D is D$_{abd}$ \cite{Ortaggio09}.} Moreover, $\bl'=\ell^\mu\pa_\mu-\ell^z\pa_z$ is also a double WAND, so that the Weyl type of $\d s^2$ is necessarily D {(in fact, D$_{abd}$, since eq.~(\ref{2WAND}) can be lifted to a timelike vector of the full space; this is nontrivial only for $n\ge 6$ \cite{Ortaggio09})}. {We can thus also conclude that when $\d s^2$ is of type D, $\d\tilde s^2$ can be (only) of the types G or I$_i$ (or D, of course).}  In order to see an explicit example, let us take $\d\tilde s^2$ to be the static vacuum KK bubble (i.e., the direct product of a timelike direction $t$ with the Euclidean Schwarzschild metric). This is a type~G spacetime \cite{GodRea09}. However, it is obvious that in the full spacetime~(\ref{ansatz}) the two null directions $\bl_{\pm}=\pa_t\pm f\pa_z$ define two distinct double WANDs ($C_{tabc}=0=C_{zabc}$ implies $C_{abcd}\ell^d_{\pm}=0$, then use \cite{Ortaggio09}) and the Weyl tensor $C_{abcd}$ is thus of type D (D$_{abd}$).

Finally, if $\bl$ is a {\em triple} (or {\em quadruple)} WAND in $\d s^2$ (with $\ell^z\neq 0$),  one can use an argument like in the previous paragraph to arrive at $\bl'$ being another triple (or quadruple) WAND, which implies the vanishing of the Weyl tensor. Non-trivial cases thus require $\ell^z=0$, so that $\d s^2$ is of type III/N if and only if $\d\tilde s^2$ is such (the ``if'' implication was discussed  in Section~\ref{sec_sharedwand}).
To summarize,  {\em if the Weyl tensor of $\d\tilde s^2$ is of type G or I, the Weyl tensor of $\d s^2$ may be more special} (in special cases). {See Table~\ref{tab_types} for details and for a summary of the whole discussion.}

\begin{table}[t]

\begin{minipage}[b]{0.5\linewidth}

\centering

\begin{center}
	\begin{tabular}{c|c}
			 Given type & Possible type \\
			 of $\d\tilde s^2$ & of $\d s^2$ \\
			 &\\[-3mm]
			 \hline &\\[-10pt]
   G	& G, I$_i$, D$_{abd}$    \\ [1pt]
		\hline &\\[-10pt]
		I & I, I$_i$  \\[1pt]
		\hline &\\[-10pt]
	  I$_i$ & I$_i$, D$_{abd}$  \\[1pt]
		\hline &\\[-10pt]
		II & II  \\[1pt]
		\hline &\\[-10pt]
		D & D  \\[1pt]
		\hline &\\[-10pt]		
		III & III \\[1pt]
		\hline &\\[-10pt]
		N & N \\[1pt]
		\hline &\\[-10pt]
		O & O \\[1pt]
		\end{tabular}
		\end{center}

\end{minipage}
\begin{minipage}[b]{0.1\linewidth}

\centering

\begin{center}
	\begin{tabular}{c|c}
			 Given type  & Possible type \\
			 of $\d s^2$ & of $\d\tilde s^2$ \\
			 &\\[-3mm]
			 \hline &\\[-10pt]
   G	& G   \\[1pt]
		\hline &\\[-10pt]
		I & I  \\[1pt]
		\hline &\\[-10pt]
		I$_i$ & G, I, I$_i$  \\[1pt]
		\hline &\\[-10pt]
		II & II  \\[1pt]
		\hline &\\[-10pt]
		D & G, I$_i$, D \\[1pt]
				\hline &\\[-10pt]
		III & III \\[1pt]
		\hline &\\[-10pt]
		N & N \\[1pt]
		\hline &\\[-10pt]
		O & O \\[1pt]		
		\end{tabular}
		\end{center}
	
\end{minipage}

\caption{Possible relation between the Weyl type of the seed spacetime $\d\tilde s^2$ and the full spacetime $\d s^2$.
}

\label{tab_types}

\end{table}

\section{Scalar invariants and curvature singularities}

\label{subsec_invariants}

\subsection{``Generic'' case}

From~(\ref{weyl}) one also readily gets for the simplest Weyl scalar 
\be
 C_{abcd}C^{abcd}=f^{-2}\tilde C_{\mu\nu\rho\sigma}\tilde C^{\mu\nu\rho\sigma} .
 \label{invariant0}
\ee

This invariant is typically nonzero for, e.g., black hole spacetimes, where it can be used to localize a curvature singularity. It is evident that such kind of singularity, when present in the seed metric, will also affect the full geometry. In addition, the latter will be singular also at zeros of $f(z)$, which are always present except in the cases $\tilde R<0$ {(for which, necessarily, $R<0$)} {and $\tilde R=0=R$ (i.e., $f_{,z}=0$)}. These additional singularities {(already previously discussed in some special cases \cite{EmpJohMye99,ChaHawRea00,LuPop01,CveLuPop01,HirKan01})} will typically extend through and beyond the (possible) event horizon, since they do not depend on the coordinates of $\d\tilde s^2$.

However, there exist spacetimes for which all invariants of zero order in the Weyl (and Riemann) tensor, such as~(\ref{invariant0}), vanish identically (VSI$_0$ spacetimes \cite{Coleyetal04vsi}) or are constant (CSI$_0$ spacetimes \cite{ColHerPel06}), and therefore cannot  be used to localize curvature singularities. {In particular, all Einstein spacetimes of type III and N fall in this class.}\footnote{ This can be easily seen as follows. For vacuum ($\Lambda=0$) type III/N spacetimes, one cannot construct non-vanishing invariants by contractions since all Weyl components have negative b.w. (whereas an invariant must necessarily be of b.w.~zero) and such spacetimes are therefore VSI$_0$. Similarly, for Einstein spacetimes of type III/N, the Weyl tensor has only negative b.w. components, while the Ricci tensor has only zero b.w. components, so that non-vanishing invariant contractions (which can thus contain only Ricci components) are clearly constant. It is also worth emphasizing that while in vacuum all type III (or more special) spacetimes are VSI$_0$ {\em and viceversa} \cite{Coleyetal04vsi}, an analogous converse implication does not hold for CSI$_0$ Einstein spacetimes. Namely, although all type III (or more special) Einstein spacetimes are CSI$_0$ (as explained above), 
there exist CSI$_0$ Einstein spacetimes that are not of such Weyl types (for example, already in four dimensions, the Nariai space \cite{Kasner25,Nariai51}, which is of type D).} In the case of expanding type III and N spacetime one can nevertheless construct certain nonzero invariants from the covariant derivatives of the Weyl tensor, as we now discuss.

\subsection{Type N and III spacetimes with nonzero expansion}

Nonzero scalar invariants for expanding spacetimes of type N/III were constructed in four dimensions in \cite{BicPra98,Pravda99} and extended to any higher dimensions in \cite{Coleyetal04vsi} (see also \cite{OrtPraPra10}). The derivation of the latter result greatly benefits from  the fact  that in type N/III Einstein spacetimes the (unique) multiple WAND $\bl$ is geodetic \cite{Pravdaetal04}. This enables one to make computations in a frame parallelly transported along $\bl$ and leads to considerable simplifications.

For {expanding} type N Einstein spacetimes one of the simplest nonzero curvature invariant is
\be
 I_{N} \equiv C^{a_1 b_1 a_2 b_2  ; c_1 c_2} C_{a_1 d_1 a_2  d_2 ; c_1 c_2} C^{e_1 d_1 e_2 d_2 ;f_1 f_2} C_{e_1 b_1 e_2 b_2  ; f_1 f_2} .
 \label{InvN}
\ee
Using a parallelly propagated frame,\footnote{For type N expanding Einstein spaces, the rank of the optical matrix $L_{ij}$ as well as the rank of the matrix of Weyl tensor components $\Psi_{ij}$ is 2 \cite{Pravdaetal04,Durkeeetal10}, {and in a suitable frame} the only non-vanishing components of $L_{ij}$ are $L_{22}=L_{33}=s$ and $L_{23}=-L_{32}=A$ {(related to~(\ref{scalars}) by $\theta=2s/(n-2)$ and $\omega=\sqrt{2}A$)}.  Note, however, that the canonical form \cite{Pravdaetal04} of the type N Weyl
tensor with the only non-vanishing component being $\Psi_{22}=-\Psi_{33}$ is not in general compatible with parallel transport, and thus in a parallelly transported frame one also needs to take into account the component  $\Psi_{23}$ \cite{OrtPraPra10}.} this invariant was shown \cite{Coleyetal04vsi} to be proportional (via a numerical constant) to 
\be
I_N\propto\left[ (\Psi_{22})^2 + (\Psi_{23})^2  \right]^2  (s^2+A^2)^4.
\ee

For the class of spacetimes~(\ref{ansatz}) considered in this paper, {from Table~\ref{tab_types} and (\ref{scalars}) it follows that $\d s^2$ is of type N with an expanding multiple WAND $\bl$ if and only if $\d\tilde s^2$ is such, and $\bl$ is simply the lifted counterpart of $\tbl$. Therefore also $\d\tilde s^2$ will admit a nonzero invariant ${\tilde I}_{N}$ defined as in~(\ref{InvN}) (similar comments will hold in the case of type III below and will not be repeated there).} One can thus substitute the Weyl components and the optical scalars given in Appendix~\ref{app_comp} to obtain
\be
I_N =\frac{1}{f^8}{\tilde I}_{N} .
\ee
This invariant was computed for certain explicit type N solutions in \cite{OrtPraPra10}.

Similarly, for {expanding} Einstein spacetimes of type III, the curvature invariant $I_{{III}}$,
\be
I_{{III}} \equiv C^{a_1 b_1 a_2 b_2;e_1} C_{a_1 c_1 a_2 c_2;e_1} C^{d_1 c_1 d_2 c_2;e_2} C_{d_1 b_1 d_2 b_2;e_2} ,
 \label{invIII}
\ee
can be expressed \cite{Coleyetal04vsi} (using the notation of \cite{OrtPraPra10}) as\footnote{This result was obtained {  using a split of $L_{ij}$ into a non-vanishing 2-block and a remaining vanishing block, which is compatible with parallel transport of the frame (as can be easily seen from \cite{OrtPraPra10}). Moreover,} an assumption on the Weyl tensor {was made} (see \cite{Coleyetal04vsi} and footnote~7 of \cite{OrtPraPra10} for details), implying that in the generic case the rank of the optical matrix $L_{ij}$ is 2. In \cite{Pravdaetal04} this was proven {in full generality in five dimensions and} in the non-twisting case in arbitrary dimension, as well as in a ``generic'' twisting higher dimensional case.  It is thus at present unclear whether rank$(L_{ij})=2$  holds in {\em all} special cases {(but surely it does for a very large class of solutions \cite{Pravdaetal04})}. Possible {``exceptional''} cases with rank$(L_{ij}) >2$ are not discussed
in this section. Note, however, that the results of the next section on p.p. singularities apply to {\em all} cases.} 
\be
I_{{III}}  \propto (s^2+A^2)^2[9(\Psi_i\Psi_i)^2+27(\Psi_i\Psi_i)({\Psi_{w22}}{\Psi_{w22}} + \Psi_{w23}\Psi_{w23})
+28({\Psi_{w22}}{\Psi_{w22}} + \Psi_{w23}\Psi_{w23})^2] .
\label{Inv3}
\ee 
Using again results from Appendix~\ref{app_comp} we get 
\be
I_{{III}} = \frac{1}{f^6}{\tilde I}_{III} .
\ee
See again \cite{OrtPraPra10} for explicit examples.

{As discussed in relation to the invariant~(\ref{invariant0}), we therefore conclude that also in the case of type N/III spacetimes with nonzero expansion points where $f(z)=0$ correspond to some curvature invariants becoming infinite.}\footnote{It is worth observing that diverging higher order invariants do not necessarily imply the presence of a ``physical'' singularity \cite{MusLak95,KonHel10}. The results of the next section however demonstrate that, at least for the spacetimes studied in this paper, points where higher order invariants become infinite are really singular (but not necessarily the other way around).} On the contrary, for type III/N Einstein spacetimes of the Kundt class (and thus with vanishing expansion) all scalar invariants constructed from the Riemann tensor and its covariant derivatives are either zero or constant \cite{Coleyetal04vsi,ColHerPel06} (see also some comments in \cite{OrtPraPra10}) and thus one cannot use them to discuss possible singularities. {Nevertheless, for the class of warped spaces~(\ref{ansatz}), one can still detect the presence of singularities by studying the frame components of the Weyl tensor as measured by a freely falling observer, which is the subject of the next section.}

\section{Freely falling observers and p.p. singularities}

\label{sec_ppsing}

In this section we introduce a class of geodetic observers and study the Weyl components measured in their frames in order to discuss spacetime singularities.

\subsection{A class of freely falling observers}

Let us assume we are given a geodetic observer in the spacetime $\d\tilde s^2$ (with coordinates $x^\mu$). This can be characterized by a unit timelike geodetic vector field $\tbt$, accompanied by $n-2$ orthonormal spacelike vectors $\tbm_{(A)}$ (with $A,B,\ldots=1,\ldots n-2$). The freely falling observer transports parallelly the frame vectors along the integral curves of $\tbt$, using an affine parameter, so that
\be
 \tilde t_{\mu||\nu}\tilde t^\nu=0 , \qquad \tilde m_{(A)\mu||\nu}\tilde t^\nu=0 ,
\ee
where a lower double bar denotes a covariant derivative in the $(n-1)$-dimensional geometry of $\d\tilde s^2$.

One can now define an observer in the full space $\d s^2$ with coordinates $x^a=(x^\mu ,z)$ by simply ``extending'' the above observer to the unit timelike $n$-dimensional vector field $\bt=f^{-1/2}\tilde t^\mu\pa_\mu$. For the remaining frame vectors one can then take $\bm_{(A)}=f^{-1/2}\tilde m^\mu_{(A)}\pa_\mu$ and $\bm_{(Z)}=\sqrt{f}\pa_z$. 
However, with the results of Appendix~\ref{app_riemann} it is easy to see that such  $\bt$ is not geodetic, except when $f_{,z}=0$.\footnote{More generally, a generic vector field ${\mbox{\boldmath{${v}$}}}=v^a\pa_a$ with $v^z=0$ cannot be geodetic unless it is null, in which case it simply corresponds to null geodesics of the slice $\d\tilde s^2$ (this was noted in \cite{ChaHawRea00,LuPop01} for special metrics belonging to the family (\ref{ansatz})).} We thus define a new $n$-dimensional frame $\{\bT,\bZ,\bm_{A}\}$ by performing a Lorentz boost in the plane of $\bt$ and $\bm_{(Z)}$, i.e.,
\be
 \bT=(\cosh\gamma)\frac{\hbt}{\sqrt{f}}+(\sinh\gamma)\sqrt{f}\pa_z, \qquad \bZ=(\sinh\gamma)\frac{\hbt}{\sqrt{f}}+(\cosh\gamma)\sqrt{f}\pa_z ,
\ee
where $\hbt=\tilde t^\mu\pa_\mu$ denotes the $n$-dimensional lift of the $(n-1)$-dimensional vector $\tbt$.

We now want to choose the function $\gamma$ such that $\bT$ is geodetic. One easily finds the condition 
$2(\tilde t^\mu\gamma_{,\mu})\cosh\gamma+f_{,z}\cosh\gamma +2\gamma_{,z}f\sinh\gamma=0$. By taking as one of the coordinates an affine parameter $\tau$ along $\tbt$, i.e.,
\be
 \tbt=\pa_\tau ,
\ee
the above condition on $\gamma$ simply reads
\be
 2\gamma_{,\tau}\cosh\gamma+f_{,z}\cosh\gamma +2\gamma_{,z}f\sinh\gamma=0 . 
 \label{geod_T}
\ee
A particularly simple solution to this equation is given, for example, by $\gamma=\gamma(z)$, which requires $\cosh\gamma=f^{-1/2}\gamma_0$ (and therefore $\sinh\gamma=\pm\sqrt{(\gamma_0^2-f)/f}$), where $\gamma_0$ is a constant (which, at least in the spacetime region of interest, must obey $\gamma_0\ge\sqrt f$). 

It follows immediately that when (\ref{geod_T}) is satisfied not only is $\bT$ geodetic but the full frame $\{\bT,\bZ,\bm_{(A)}\}$ is in fact parallelly transported along $\bT$, so that we have constructed a freely falling observer in $\d s^2$.

\subsection{Frame components of the Weyl tensor and singularities}

\label{subsec_sing_time}

Recalling~(\ref{weyl}) we can now relate the frame components of the Weyl tensor $C_{abcd}$ measured by such an observer to those of the Weyl tensor $\tilde C_{\mu\nu\rho\sigma}$ measured by the $(n-1)$-dimensional observer $\{\tbt,\tbm_{(A)}\}$. One readily finds
\beqn
 & & C_{TATB}=f^{-1}\cosh^2\gamma\tilde C_{ t A t B} , \quad C_{TAZB}=f^{-1}\sinh\gamma\cosh\gamma\tilde C_{ t A t B} , \quad  C_{ZAZB}=f^{-1}\sinh^2\gamma\tilde C_{ t A t B} , \nonumber \label{weyt_time} \\
 & & C_{TABC}=f^{-1}\cosh\gamma\tilde C_{ t A B C} , \quad C_{ZABC}=f^{-1}\sinh\gamma\tilde C_{ t A B C} , \quad 
  C_{ABCD}=f^{-1}\tilde C_{ A B C D} .
\eeqn
(In the above equations, lower indices $T$ and $t$ denote contraction with, respectively, $\bT$ and $\tbt$.)
 While in each Weyl component there appear an explicit $z$ dependence through the factor $f^{-1}$, we observe that, due to (\ref{geod_T}), also the hyperbolic functions necessarily depend on~$z$ (except in the case $f_{,z}=0$, corresponding to a direct product spacetime). In particular, for the special solution mentioned above, 
\be
 \cosh\gamma=f^{-1/2}\gamma_0 ,
\ee
each $k$-power of a hyperbolic function will introduce an extra $f^{-k/2}$ factor in the corresponding Weyl component. Therefore, all the frame components of the Weyl tensor become singular at zeros of $f(z)$

We have thus demonstrated that for any geodetic observer $\tilde t$ in the seed geometry $\d\tilde s^2$, there exists (at least) one geodetic observer $\bT$ in the full geometry $\d s^2$ for which the Weyl components, as measured in a parallelly transported frame, blow up at points where $f(z)=0$. This is true independently of possible singularities of the spacetime $\d\tilde s^2$, and therefore even if $\d\tilde s^2$ is assumed to be everywhere regular (unless it is conformally flat). However, we have to show that $\bT$ really encounters some points with $f(z)=0$. This can be seen as follows. We can define an affine parameter $\tilde z$ along the vector field $\bT=f^{-1}\gamma_0\pa_\tau\pm\sqrt{\gamma_0^2-f}\pa_z$ by taking 
\[
 \d\tilde z=\pm(\gamma_0^2-f)^{-1/2}\d z , 
\] 
so that $\bT=f^{-1}\gamma_0\pa_\tau+\pa_{\tilde z}$. The affine parameter $\tilde z$ is thus monotonically increasing [decreasing] along the coordinate $z$.  
The explicit form of $\tilde z(z)$ depends on the sign of $\lambda$. Nevertheless, in all cases one can see that a zero of $f(z)$ (if present) will be reached at a finite value of the affine parameter $\tilde z$.\footnote{The only exception to this may possibly occur in the case $\bT$ reaches (at a special value of $\tau$) a singularity inherited from the seed metric $\d\tilde s^2$ before reaching $f(z)=0$. However, this will generically not be the case. See, e.g., \cite{ChaHawRea00,LuPop01} for explicit examples.} In other words, we have shown that {\em the spacetime $\d s^2$ possesses a p.p. curvature singularity} \cite{HawEll73} at points where $f(z)=0$. Not surprisingly, these are the same points where the scalar invariants discussed in the previous sections diverge. However, p.p. curvature singularities will generically be present also in spacetimes where invariants do not diverge (either because they are identically zero or constant, such as in Kundt solutions of type N and III, or because, although not constant, they still remain finite along certain geodesics, cf.~\cite{ChaHawRea00,LuPop01}). Recall, however, that such singularities do not occur in the $(-,-)$ (i.e., when $R<0$ and $\tilde R<0$) and in the $(0,0)$ cases, for which $f(z)$ never vanishes.


\section{A few examples}

\label{sec_examples}

In previous sections we have already mentioned certain explicit spacetimes that have been constructed using the ansatz~(\ref{ansatz}) considered in this paper. In particular, various static black string solutions have been studied \cite{ChaHawRea00,HirKan01,GodRea09}. By the results of Section~\ref{sec_weyl}  these are all of type D (this follows also from the more general results of \cite{PraPraOrt07}, see also \cite{GodRea09}). 

It is now straightforward to extend all such strings to the spinning case by simply taking a spinning black hole as a seed metric. As we do not want to have naked singularities in the full spacetime, we restrict to the case of an AdS string sliced by an AdS black hole. For example, we can take $\d\tilde s^2$ to be the four-dimensional Kerr-AdS solution \cite{Carter68cmp}, so that we get the five-dimensional rotating black string (cf. metric~(\ref{--b}))
\beqn
 \d s^2= & & \left[\d x^2-\frac{\Delta_r}{\Xi^2\rho^2}\left(\d t-a\sin^2\theta\d\phi\right)^2+\frac{\Delta_\theta\sin^2\theta}{\Xi^2\rho^2}\left(a\d t-(r^2+a^2)\d\phi\right)^2+\rho^2\left(\frac{\d r^2}{\Delta_r}+\frac{\d\theta^2}{\Delta_\theta}\right)\right] \nonumber \\
 & & {}\times\cos^{-2}(\sqrt{-\lambda}x) ,
\eeqn
where $\lambda<0$, $\sqrt{-\lambda}x\in(-\pi/2,\pi/2)$, and
\beqn
 & & \rho^2=r^2+a^2\cos^2\theta, \qquad \Xi=1+\lambda a^2 , \nonumber \\
 & & \Delta_r=(r^2+a^2)(1-\lambda r^2)-2mr , \qquad \Delta_\theta=1+\lambda a^2\cos^2\theta .
\eeqn
The most important features of the above spacetime are inherited from the four-dimensional seed, e.g., the horizon and ergosurface structure, after taking into account the warped extra dimension. Additionally, $\sqrt{-\lambda}x\to\pm\pi/2$ can be interpreted as the (bulk) AdS timelike infinity.\footnote{By this we simply mean that the spacetime $\d s^2$ admits a conformal boundary at $\Omega=0$, with $\Omega=\cos(\sqrt{-\lambda}x)$, whose normal ${\cal N}_{a}=\Omega_{,a}$ is a spacelike vector in the conformal geometry $\d\hat s^2=\Omega^2\d s^2$. In particular, we do not claim that these warped spacetimes are asymptotically AdS, and in fact they are not, in general (according to the definitions of , e.g., \cite{AshDas00,HolIshMar05}).} This metric can be of course extended to any higher dimensions by using the higher dimensional rotating AdS black holes of \cite{HawHunTay99,Gibbonsetal04_jgp} as seed spacetimes. {All such solutions are of type D (cf. Table~\ref{tab_types}).}

Further, using the four-dimensional AdS C-metric \cite{PleDem76} as a seed, one gets a five-dimensional accelerating black string, i.e.,
\begin{equation} 
  \d s^2 =\cos^{-2}(\sqrt{-\lambda}x)\left[\d x^2+\frac1{\accl^2(w+\yKW)^2}\left(-\FKW(y) \d\tKW^2 +\frac{\d y^2}{\FKW(y)}+\frac{\d w^2}{\GKW(w)} +\GKW(w) \d\phi^2\right)\right] ,
\end{equation}
where 
\begin{equation}
  \FKW(y)=-\frac{\lambda}{\accl^2}-1+y^2-2\mass\accl y^3 , \qquad \GKW(w)=1-w^2-2\mass\accl w^3 .
\end{equation}
Most of the properties of this solution (e.g., according to the value of the acceleration parameter, its maximal analytical extension can represent one or two accelerating black strings) follow from the results known in four dimensions, see, e.g., \cite{Krtous05} and references therein. {Again, this is a type D solution.}
At present no exact solution analogue to the C-metric in more than four dimensions is known, therefore accelerating strings can be constructed only for $n=5$, with this method.

One can also combine the above solutions and use the spinning AdS C-metric \cite{PleDem76} (see also \cite{GriPodbook} and references therein) to construct black strings {(of type D)} in five dimensions that are both accelerating and rotating. The method is now obvious and we shall not write down the resulting metric explicitly.

\section{Concluding remarks}

\label{sec_conclusions}

We have analyzed various specific properties of a class of higher dimensional Einstein spacetimes, which are naturally singled out in the theory of conformal Einstein spaces \cite{Brinkmann25}. Several instances of such metrics had appeared previously in the literature in different contexts. We have however related various such coordinate representations (using the unified coordinates of Brinkmann) in a systematic way and we have analyzed geometric properties (Weyl type, curvature invariants, singularities) that characterize these spacetimes and which are important from the viewpoint of the recently developed NP formalism. {It is also worth emphasizing that} while in most cases naked singularities appear due to the warped product, interestingly this is not so when the cosmological constants of both metrics $\d\tilde s^2$ and $\d s^2$ are negative.
We have also emphasized how Brinkmann's metric can be used as a useful ansatz to generate new solutions of possible interest. In particular, some explicit examples representing certain black strings have also been provided. We observe that the same warped metrics have already been considered in theories different form pure Einstein gravity, e.g., in the braneworld KK reductions studied in \cite{LuPop01,CveLuPop01,ParPopSad02}. There would not be any obstacles in straightforwardly extending most of our analysis to such theories. On the other hand, it would be interesting to see how the results of the present paper could be generalized to more general warped spacetimes, which is left for possible future work.

\section*{Acknowledgments}

This work has been supported by {research plan No AV0Z10190503 and research grant GA\v CR \\
P203/10/0749}.

\appendix

\section{Christoffel symbols, Riemann and Ricci tensors}

\label{app_riemann}

\renewcommand{\theequation}{A\arabic{equation}}
\setcounter{equation}{0}

For completeness let us present here the Christoffel symbols, the Riemann and Ricci tensors for the metric~(\ref{ansatz}), which are used in the paper. Most of these relations can also be found in \cite{Brinkmann25,petrov}.

First, from (\ref{ansatz}) we obviously have
\be
g_{\mu\nu}=f(z){\tilde g}_{\mu\nu}, \qquad g_{zz}=\frac{1}{f(z)}.
\ee
The Christoffel symbols read
\bea
\Gamma^z_{\mu\nu}&=&-\frac{ff_{,z}}{2}{\tilde g}_{\mu\nu}, \qquad 
\Gamma^z_{z\mu }=0, \qquad 
\Gamma^z_{zz}=-\frac{f_{,z}}{2f},\\
\Gamma^\mu _{\nu\rho}&=&{\tilde \Gamma}^\mu _{\nu\rho}, \qquad 
\Gamma^\mu _{\nu z}=\frac{f_{,z}}{2f}{\delta}^\mu _\nu , \qquad 
\Gamma^\mu _{zz}=0 .
\eea

For the Riemann tensor one finds
\bea
{R^z}_{\mu  z\nu }&=&-\frac{ff_{,zz}}{2}{\tilde g}_{\mu\nu}, \qquad
{R^z}_{\mu\nu\rho}=0,\ \\
{R^\mu }_{\nu \rho\sigma}&=&{{\tilde{R}}^\mu }_{\ \nu\rho\sigma}+\frac{(f_{,z})^2}{4}(\delta^\mu _\sigma {\tilde g}_{\nu\rho}-\delta^\mu _\rho {\tilde g}_{\nu\sigma}),
\eea
so that the Ricci tensor is 
\bea
R_{\mu\nu}&=&{R^d}_{\mu  d\nu }={\tilde R}_{\mu\nu}-\left[\frac{ff_{,zz}}{2}+\frac{n-2}{4}(f_{,z})^2\right] {\tilde g}_{\mu\nu},\\
R_{zz}&=&-\frac{(n-1)f_{,zz}}{2f} , \qquad R_{\mu  z}=0.
\eea
It is then easy to see that 
\be
R=\frac{\tilde R}{f}-(n-1)\left[ f_{,zz}+\frac{(n-2)}{4f}(f_{,z})^2\right].
\ee

Let us now assume that the spacetime $\d s^2$ is Einstein, as in the main text. The Einstein equations 
then give 
\be
R_{\mu\nu}=(n-1)\lambda g_{\mu\nu}=\frac{R}{n}g_{\mu\nu}, \qquad R_{zz}=\frac{R}{n}g_{zz}, \qquad  R_{\mu  z}=0 
\label{EinsteinEqs} ,
\ee
{where $\lambda$ is a constant.} From these one finds
\bea
R&=&n(n-1)\lambda,\\
f,_{zz}&=&-2\lambda=\mbox{const}\ \ \Rightarrow\ \ f=-\lambda z^2+2dz+b,\\
{\tilde R}_{\mu\nu}&=&
\frac{{\tilde R}}{n-1}{\tilde g}_{\mu\nu}, \qquad   {\tilde R}=(n-1)(n-2)(\lambda b+d^2) ,
\eea
i.e. the seed metric is automatically also Einstein.
Using the above equations, from the definition of the Weyl tensor one can also obtain the result~(\ref{weyl}) for the Weyl tensor of Einstein spaces.

\section{Optical matrix and Weyl tensor components in a frame parallelly propagated along a geodesic $\bl$}

\label{app_comp}

\renewcommand{\theequation}{B\arabic{equation}}
\setcounter{equation}{0}

In this appendix we define a family of null frames that are parallelly transported along a geodetic vector field $\bl$ and evaluate the corresponding Weyl tensor components.

In section~\ref{sec_weyl} we employed a null frame obtained by simply lifting a null frame of $\d\tilde s^2$. 
In particular, when $\tbl$ is geodetic and affinely parametrized, $\bl$ (as defined in (\ref{lifted_frame})) inherits the same property \cite{PraPraOrt07} ($\bl$ considered in \cite{PraPraOrt07} corresponds to $f^{-1}$ times the $\bl$ of the present paper, but this does not affect the previous statement) and we can take as one of our coordinates an affine parameter $r$, so that $\bl=\pa_r$. In such a case, one can define a frame which is parallelly propagated along $\bl$, which may be useful for various  purposes (in particular, to express the Weyl tensor components and describe their possible peeling-off properties, cf.~\cite{OrtPraPra10}). However, if one naturally starts from a frame  $\{ \tbl ,\ \tbn ,\ \tbm_{(J)} \}$  in $\tilde\d s^2$ that is parallelly propagated along $\tbl $, the frame vectors $\bn$ and $\bm_{(Z)}$ defined in (\ref{lifted_frame}) will {\em not} be parallelly transported along $\bl$ (except when $f_{,z}=0$). 
A parallelly transported frame can however be obtained by performing the following null rotation of (\ref{lifted_frame})  
\BE
 \bl\to\bl, \qquad \bn\to\bn+\zeta\bm_{(Z)} -\pul \zeta^2\bl , \qquad \bm_{(I)}\to\bm_{(I)} , \quad \bm_{(Z)}\to\bm_{(Z)}-\zeta\bl ,
 \label{nullrot}
\EE
with $\zeta=\frac{1}{2}f^{-1/2}f_{,z}r$  (up to an arbitrary additive term independent of $r$).

The new, parallelly transported frame thus reads\footnote{To verify this one needs the relations 
$ {\tilde g}_{rr}=0$ , ${\tilde\Gamma}^\mu _{rr}=0$, $\Gamma^z_{rr}=0$ , ${\tilde g}_{\mu  r}{\tilde m}^\mu _{(J)}
=0$, ${\tilde g}_{\mu  r}m^\mu _{(Z)}
=0$, ${\tilde n}_r=1$, 
which follow from the orthonormality conditions on the parallelly transported frame  $\{ \tbl ,\ \tbn ,\ \tbm_{(J)} \}$ and the results of Appendix~\ref{app_riemann} rewritten in the coordinates of this section.}
\beqn
 & & \bl=\pa_r, \qquad \bn=\frac{1}{f}\tilde n^\mu\pa_\mu+\frac{rf_{,z}}{2}\pa_z-\frac{r^2(f_{,z})^2}{8f}\pa_r, \nonumber \\
 & & \bm_{(I)}=\frac{1}{f^{1/2}}\tilde m^\mu_{(I)}\pa_\mu, \qquad \bm_{(Z)}=\sqrt{f}\pa_z-\frac{rf_{,z}}{2f^{1/2}}\pa_r . \label{rotated_frame}
\eeqn

Now we can compute the optical matrix and the Weyl frame components in this frame and compare them with those of the seed geometry $\d\tilde s^2$.  
For the optical matrix and the optical scalars we obtain\footnote{Note that, since $\bl$ is geodetic, the optical matrix $L_{ij}$ (\ref{L}) and the optical scalars (\ref{scalars}) are invariant under null rotations \cite{OrtPraPra07}. One can thus compute these in the frame (\ref{lifted_frame}) and the result holds unchanged also in the frame~(\ref{rotated_frame}).}
\bea
L_{JK}&=&{\tilde L}_{JK}, \qquad
L_{JZ}=0=L_{ZJ}, \qquad 
L_{ZZ}=0, \label{L} \\
 \sigma^2&=&{\tilde \sigma}^2+\frac{n-3}{n-2}{\tilde \theta}^2, \qquad \theta=\frac{n-3}{n-2}{\tilde \theta}, \qquad \omega^2={\tilde \omega}^2 . 
 \label{scalars}
\eea
(Cf.~an equivalent result in eq.~(18) of \cite{PraPraOrt07}.) While expansion and twist are essentially the same in the seed and in the full geometry, the presence of expansion in the seed geometry gives rise to shear in the full geometry, even if $\tbl$ is shearfree. 

The Weyl components in the frame~(\ref{rotated_frame}) have the form
\bea
\mbox{Boost weight $+2$:}&& \nonumber\\
C_{0J0K}&=&{\tilde C}_{0J0K},\qquad 
C_{0Z0Z}=0,\qquad 
C_{0Z0J}=0.\\
\mbox{Boost weight $+1$:}&& \nonumber\\
C_{010J}&=&\frac{1}{\sqrt{f}}{\tilde C}_{010J},\qquad 
C_{010Z}=0, \qquad 
C_{0JKL}=\frac{1}{\sqrt{f}}{\tilde C}_{0JKL},\ \ \nonumber\\
C_{0ZJK}&=&0=C_{0ZJZ},\qquad 
C_{0JZK}=-\frac{f_{,z} r}{2\sqrt{f}}{\tilde C}_{0J0K}.\\
\mbox{Boost weight  0:}&&\nonumber\\
C_{01JK}&=&\frac{1}{f}{\tilde C}_{01JK},\qquad 
C_{IJKL}=\frac{1}{f}{\tilde C}_{IJKL}, \qquad  C_{0101}=\frac{1}{f}{\tilde C}_{0101},\nonumber\\
C_{0J1K}&=&\frac{1}{f}{\tilde C}_{0J1K}+N {\tilde C}_{0J0K},\qquad 
C_{01JZ}=\frac{f_{,z} r}{2f}{\tilde C}_{010J},\ \ \nonumber\\
C_{0J1Z}&=&\frac{f_{,z} r}{2f}{\tilde C}_{010J},\qquad 
C_{0Z1J}=0=C_{0Z1Z},\nonumber\\
C_{ZJKL}&=&-\frac{f_{,z} r}{2f}{\tilde C}_{0JKL},\qquad 
C_{ZJZK}=-2N{\tilde C}_{0J0K}.\ \ \\
\mbox{Boost weight $-1$:}&&\nonumber\\
C_{1JKL}&=&\frac{1}{f^{3/2}}{\tilde C}_{1JKL}+\frac{1}{f^{1/2}}{\tilde C}_{0JKL}N ,\qquad 
C_{1ZJK}=\frac{f_{,z} r}{2f^{3/2}}{\tilde C}_{01JK},\ \ \nonumber\\
C_{1JZK}&=&-\frac{f_{,z} r}{2f^{3/2}}{\tilde C}_{0K1J}-N  \frac{f_{,z} r}{2\sqrt{f}}{\tilde C}_{0J0K},\qquad 
C_{1ZJZ}=\frac{(f_{,z} r)^2}{4f^{3/2}}{\tilde C}_{010J},\ \ \nonumber\\
C_{101J}&=&\frac{1}{f^{3/2}}{\tilde C}_{101J}-\frac{1}{f^{1/2}}{\tilde C}_{010J}N ,\qquad 
C_{101Z}=-\frac{f_{,z} r}{2f^{3/2}}{\tilde C}_{0101}.\ \ \\
\mbox{Boost weight $-2$:}&&\nonumber\\
C_{1J1K}&=&\frac{1}{f^{2}}{\tilde C}_{1J1K}+\frac{N }{f}({\tilde C}_{0J1K}+{\tilde C}_{0K1J})
+{\tilde C}_{0J0K}(N )^2,\ \ \nonumber\\
C_{1Z1J}&=&-\frac{f_{,z} r}{2f^2}{\tilde C}_{101J}+\frac{f_{,z} r}{2f}N  {\tilde C}_{010J},\qquad 
C_{1Z1Z}=\frac{(f_{,z} r)^2}{4f^2}{\tilde C}_{0101}, \ \
\eea
where $N=-\frac{(f_{,z}r)^2}{8f}$.

{By knowing the $r$-dependence of the Weyl tensor of a seed spacetime, the above results enable one to characterize the behavior of the Weyl tensor of the full space $\d s^2$. This can be used, for example, for discussing peeling properties as one moves along the null direction $\bl$ (see \cite{OrtPraPra10} for an explicit analysis in the case of type N/III spacetimes).}


\end{document}